\documentclass[twocolumn,preprintnumbers, amssymb,amsmath,aps,floatfix,nofootinbib,superscriptaddress,showpacs]{revtex4-1}

\usepackage{epsfig}
\usepackage{bm}
\usepackage{amssymb}
\usepackage{amsmath}
\usepackage{color}
\usepackage{subfigure}
\usepackage[colorlinks,
            linkcolor=blue,
            anchorcolor=black,
            citecolor=blue
            ]{hyperref}

\allowdisplaybreaks[4]

\newcommand\Q{\mathcal{Q}}
\newcommand\Qb{\bar{\mathcal{Q}}}

\makeatother

\begin{document}
\title{The Collectivity of Heavy Mesons in Proton-Nucleus Collisions}

\author{Cheng Zhang} 
\affiliation{Key Laboratory of Quark and Lepton Physics (MOE) and Institute of Particle Physics, Central China Normal University, Wuhan 430079, China}
\affiliation{Key Laboratory of Particle Physics and Particle Irradiation (MOE), Institute of frontier and interdisciplinary science, Shandong University, QingDao, Shandong 266237, China}

\author{Cyrille Marquet} \email{cyrille.marquet@polytechnique.edu}
\affiliation{CPHT, CNRS, Ecole Polytechnique, Institut Polytechnique de Paris, Route de Saclay, 91128 Palaiseau, France}

\author{Guang-You Qin} \email{guangyou.qin@mail.ccnu.edu.cn}
\affiliation{Key Laboratory of Quark and Lepton Physics (MOE) and Institute of Particle Physics, Central China Normal University, Wuhan 430079, China}
\affiliation{Nuclear Science Division Mailstop 70R0319,  Lawrence Berkeley National Laboratory, Berkeley, California 94740, USA}

\author{Yu Shi} 
\affiliation{Key Laboratory of Quark and Lepton Physics (MOE) and Institute of Particle Physics, Central China Normal University, Wuhan 430079, China}

\author{Lei Wang}
\affiliation{Key Laboratory of Quark and Lepton Physics (MOE) and Institute of Particle Physics, Central China Normal University, Wuhan 430079, China}

\author{Shu-Yi Wei}   \email{shu-yi.wei@polytechnique.edu}
\affiliation{European Centre for Theoretical Studies in Nuclear Physics and Related Areas (ECT*)
and Fondazione Bruno Kessler, Strada delle Tabarelle 286, I-38123 Villazzano (TN), Italy}
\affiliation{CPHT, CNRS, Ecole Polytechnique, Institut Polytechnique de Paris, Route de Saclay, 91128 Palaiseau, France}

\author{Bo-Wen Xiao}   \email{bxiao@mail.ccnu.edu.cn}
\affiliation{Key Laboratory of Quark and Lepton Physics (MOE) and Institute
of Particle Physics, Central China Normal University, Wuhan 430079, China}
\affiliation{CPHT, CNRS, Ecole Polytechnique, Institut Polytechnique de Paris, Route de Saclay, 91128 Palaiseau, France}

\begin{abstract}
Using a model based on the Color Glass Condensate framework and the dilute-dense factorization, we systematically study the azimuthal angular correlations between a heavy flavor meson and a light reference particle in proton-nucleus collisions. The obtained second harmonic coefficients (also known as the elliptic flows) for $J/\psi$ and $D^0$ agree with recent experimental data from the LHC. We also provide predictions for the elliptic flows of $\Upsilon$ and $B$ meson, which can be measured in the near future at the LHC. This work can shed light on the physics origin of the collectivity phenomenon in the collisions of small systems. 
\end{abstract}

\maketitle

\section{introduction}

In the last decade, the collectivity phenomenon in small collisional systems (such as proton-proton and proton-nucleus collisions) has been an extremely interesting topic in heavy-ion physics, with a lot of experimental evidences\cite{Khachatryan:2010gv, CMS:2012qk, Abelev:2012ola, Aad:2012gla, Adare:2013piz, Adare:2014keg, Khachatryan:2015waa, Aidala:2018mcw} observed in high multiplicity events. This particular phenomenon, which implies non-trivial angular correlations among many produced particles in small systems, can be conveniently described by the Fourier harmonics of the azimuthal angular distribution of the measured particles. If one chooses a reference particle and define $\Delta \phi$ as the azimuthal angle difference between the measured particle and the reference, one can quantitatively extract these Fourier harmonics coefficients $v_n \equiv \langle \cos n \Delta \phi \rangle$ in high-multiplicity events and find that the azimuthal angular distribution of particles is close to isotropic with a small anisotropy characterized by $v_n$ coefficients whose magnitudes are about a few percent.

More remarkably, recent measurements in $pPb$ collisions by ALICE\cite{Acharya:2017tfn} and CMS\cite{CMS:2018xac, Sirunyan:2018toe, CMS:2019isc} showed that heavy mesons such as $J/\psi$ and $D^0$ have the elliptic flow $v_2$ comparable to the $v_2$ values of light hadrons. In the near future, the elliptic flow of heavier mesons such as $B$ mesons and $\Upsilon$ may also be measured with sufficiently abundant statistics at the high-luminosity LHC.

This type of azimuthal angular correlation can have an apparent and intuitive classical interpretation in terms of pressure gradients in relativistic hydrodynamics. In the hydrodynamical approach, it is believed that a small droplet of quark-gluon plasma (QGP) is created even in the high-multiplicity events when two protons or a proton and a heavy nucleus collide. Small spatial anisotropies are generated by initial collisional geometries with small fluctuations in the high-multiplicity events in small collisional systems. The relativistic hydrodynamical evolution, which essentially conserves energy and momentum, is used to describe the space-time evolution of the QGP droplet in the final state after its initial production. Together with other final-state effects including the interaction between the probes and the QGP medium, it can convert the initial spatial anisotropy of the QGP droplet into the anisotropy in the momentum space of the measured particles in the final state. Quantitatively, hydrodynamical approaches\cite{arXiv:1304.3044,arXiv:1304.3403,arxiv:1306.3439,arXiv:1307.4379,arXiv:1307.5060,arXiv:1312.4565,arXiv:1405.3605,Habich:2014jna, arXiv:1409.2160, arXiv:1609.02590,arXiv:1701.07145,arXiv:1801.00271} have been very successful in explaining the experimental results and making predictions for the collective behaviors involving light hadrons. Nevertheless, usually due to large masses, heavy flavor particles do not flow as much as light particles. A recent study\cite{Du:2018wsj} indicates that the final-state interactions between heavy quarks and the QGP medium are not sufficient to generate enough $v_2$ to explain the observed elliptic flow for $J/\psi$ and $D^0$ mesons. Alternative mechanisms and additional sources of anisotropy besides the usual hydrodynamics approach are also proposed in Refs.\cite{Lin:2003jy,arXiv:1803.02072, Li:2018leh, Kurkela:2018qeb}.

In the meantime, the observed collectivity may also have an interesting underlying quantum interpretation from the perspective of the color glass condensate (CGC) framework\cite{Armesto:2006bv, Dumitru:2008wn, Gavin:2008ev, Dumitru:2010mv, Dumitru:2010iy, Kovner:2010xk, Kovchegov:2012nd, Dusling:2012iga, Kovchegov:2013ewa, Dumitru:2014dra, Dumitru:2014yza, Dumitru:2014vka, Lappi:2015vha, Schenke:2015aqa, Lappi:2015vta,McLerran:2016snu, Kovner:2016jfp, Iancu:2017fzn, Dusling:2017dqg, Dusling:2017aot, Fukushima:2017mko, Kovchegov:2018jun, Boer:2018vdi, Mace:2018vwq, Mace:2018yvl,Altinoluk:2018ogz,Kovner:2018fxj,Kovner:2017ssr,Kovner:2018vec, Davy:2018hsl, Zhang:2019dth}. Due to the multiple interactions during the initial collision and the quantum evolution of the dense background gluon fields in high energy $pPb$ collisions, the produced particles usually also have small correlations, which can also be interpreted as collective behavior. During the initial-state production process prior to the onset of hydrodynamical evolutions, uncorrelated active partons from the projectile proton strongly interact with the dense gluonic background fields inside the target heavy nucleus and can pick up non-trivial quantum correlations described by the so-called quadrupole correlators\cite{Dominguez:2011wm, Dominguez:2012ad} in the CGC framework. In the language of Mueller's dipole model, the quadrupole configuration involved in the correlation arises by conversion from a two-color-dipole configuration due to the so-called inelastic multiple scatterings. Therefore, the angular correlation, which is $1/N_c^2$ suppressed, can be built up from two initially independent dipoles. In particular, the collective behavior of light hadrons produced in $pA$ collisions can also be quantitatively explained within this framework\cite{Davy:2018hsl}. Moreover, the CGC framework has been demonstrated to be important in understanding the heavy quarkonium productions\cite{Ma:2014mri, Ma:2015sia, Watanabe:2015yca, Ma:2018bax} in $pp$ and $pPb$ collisions in the low transverse momentum region.

The objective of this paper is to extend our previous work\cite{Zhang:2019dth} and compute the elliptic flow $v_2$ for both quarkonia and open heavy mesons. Our result shows that initial state effects due to the strong background gluon fields inside target nucleus can generate sufficient amount of collectivity for heavy quarkonia and open heavy flavor mesons. With reasonable choices of parameters and within the range of validity of our CGC model, we find that one can explain the large $v_2$ for both $J/\psi$ and $D^0$ mesons measured by the CMS collaboration. We also make the prediction for the $v_2$ of $\Upsilon$ and $B$ mesons which can be measured in the near future.

The paper is organized as follows. In Sec.$\text{\uppercase\expandafter{\romannumeral2}}$, we provide detailed derivations for the second anisotropy harmonics of heavy quarkoium, heavy quark or open heavy meson with respect to a light quark in $pA$ collisions in a CGC model. In Sec.$\text{\uppercase\expandafter{\romannumeral3}}$, the numerical results are presented. The conclusion and further discussions are given in Sec.$\text{\uppercase\expandafter{\romannumeral4}}$.

\section{Correlations in the CGC Formalism}
To study the angular correlation of heavy mesons, let us first build a model for the production of a heavy-quark pair and a reference quark in the CGC formalism in high-multiplicity events of $pA$ collisions. In high-multiplicity $pPb$ collisions, we assume that many active partons from the proton projectile participate in the strong interaction with the target nucleus and get produced in the final state. For the purpose of studying the flow of heavy mesons, we consider the production of an incoming gluon, which splits into a pair of heavy quark and anti-quark, together with an incoming quark, which serves as a reference particle, in the presence of the strong gluonic background field generated by the target nucleus. In our model, we assume that the incoming gluon and the reference quark are initially independent, and therefore they have little correlation in both rapidity and azimuthal angle before they interact with the target nucleus. Furthermore, the dominant contribution of this process does not generate any correlation, since the incoming gluon (or the split heavy-quark pair) and the reference particle can interact with the background gluon fields independently, without knowing of the existence of the other. Nevertheless, angular correlations can be built up due to color interferences, if they start to interact with the same color charges in the nucleus simultaneously.    

As we show below, in order to obtain non-trivial correlations, we need to evaluate the expectation value of products of dipole amplitudes, up to $\frac{1}{N_c^2}$ order, in the dense background gluon fields of target nucleus in the language of the CGC framework. Furthermore, we can obtain the correlation between the final state quarkonium and the spectator quark by keeping the total momentum of the heavy-quark pair fixed while integrating over their relative momentum. This is used to calculate the elliptic flow of $J/\Psi$ and $\Upsilon$ in pA collisions. Alternatively, one may integrate over the momentum of either the heavy-quark or the anti-heavy-quark, then the collectivity of open heavy flavor mesons can be studied. As the common practice to conserve transverse momentum, we choose to work in the coordinate space which makes the summation of the multiple scattering with the dense nuclear target straightforward in the eikonal limit.

\subsection{Differential rate of the $p+A \to \Q\Qb + q + X$ process}

\begin{figure}[h!]
\includegraphics[width=0.49\textwidth]{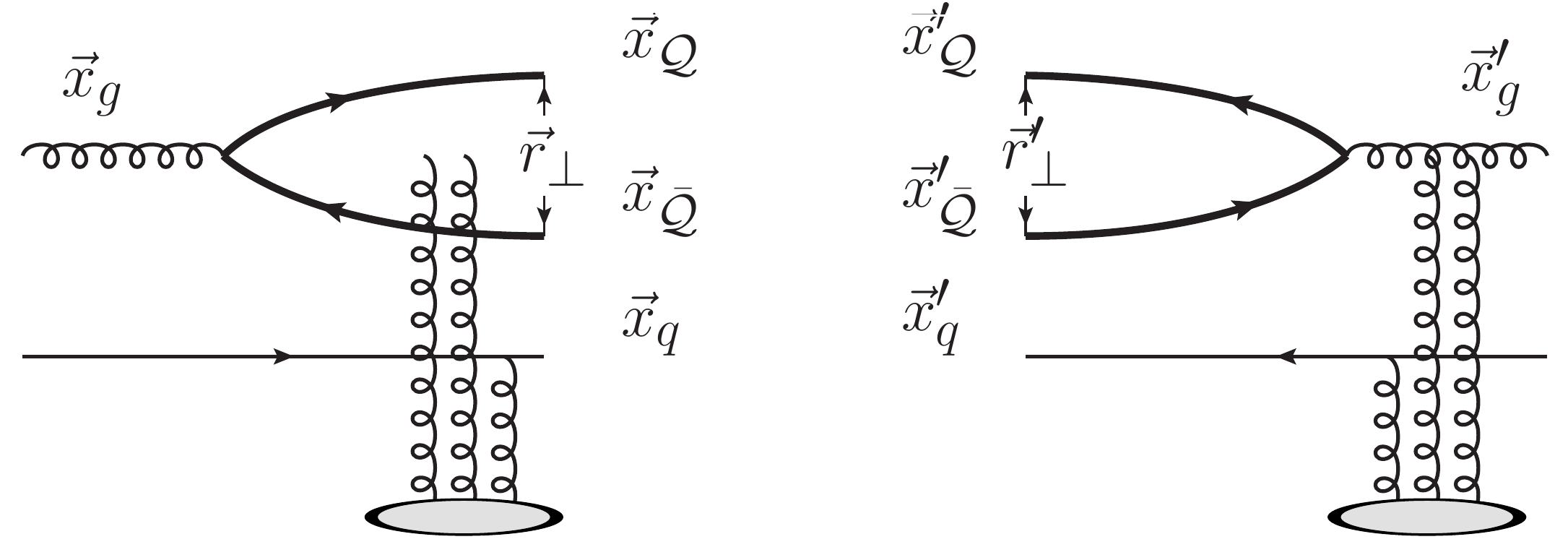}
\caption{An example of Feynman diagrams that contribute to the differential spectrum of heavy quark pair production accompanied by a spectator quark. The transverse coordinates of partons are illustrated in the diagrams. This diagram shows the interference contribution before and after the splitting of the heavy quark pair. The total contribution also includes the squares of the above individual amplitudes.}
\label{fig:feynman}
\end{figure}

In the CGC formalism, the differential spectrum for the production of a heavy quark-antiquark pair plus a spectator quark can be obtained by employing techniques developed in \cite{Gelis:2001da, Blaizot:2004wv, JalilianMarian:2004da, Dominguez:2008aa, Marquet:2010cf, Dominguez:2012ad,Shi:2017gcq,Zhang:2019yhk} and calculating diagrams as illustrated in Fig.~\ref{fig:feynman}. The resulting expression is given by
\begin{widetext}
\begin{align}
\left. \frac{dN}{d{\mathcal{PS}}} \right|_{\Q\Qb q}
= 
~& 
\mathcal{N}
\int \frac{d^2b_1 d^2b_2d^2r_1d^2 r_2 d^2r d^2r'}{(2\pi)^9} 
e^{-i\vec k_{\Q\perp} \cdot [\vec r_1+(1-\zeta)(\vec r-\vec r')]} 
e^{-i\vec k_{\Qb\perp} \cdot [\vec r_1 - \zeta (\vec r- \vec r')]} 
e^{-i\vec k_{q\perp} \cdot \vec r_2} 
\nonumber \\
& 
\times
W(x_g,x_q, b_1,b_2,r_1,r_2) 
k_g^+ \sum_{\alpha\beta\lambda}\psi_{\alpha\beta}^{T\lambda}(\vec r) \psi_{\alpha\beta}^{T\lambda*} (\vec r')
\langle DDD \rangle
,\label{eq:cs-qqq}
\end{align}
where $\mathcal{N}$ is the overall normalization factor and $\zeta = k_{\Q}^+/k_g^+$ is the longitudinal momentum fraction carried by the heavy quark.
The transverse coordinates used above are illustrated in Fig. \ref{fig:feynman} with $\vec r_1 = \vec x_g - \vec x'_g$ and $\vec r_2 = \vec x_q - \vec x'_q$. Here
$d{\mathcal{PS}} = dy_{\Q} d^2k_{\Q\perp} d\zeta d^2k_{\Qb\perp} dy_q d^2k_{q\perp}$ is the differential phase space volume of the final states. To model the double parton distribution from the proton projectile, we use
$W(x_g,x_q, b_1,b_2,r_1,r_2) = x_g f_g(x_g) x_q f_q(x_q)\frac{1}{\pi^2 B_p^2}\exp [-\frac{b_1^2+b_2^2}{B_p} - \frac{\Delta^2 (r_1^2+r_2^2)}{4}]$ which specifies the momentum and spatial distributions of the initial state gluon-quark pair in the proton, with an overall normalization factor being absorbed into $\mathcal{N}$. The parameter $B_p$ controls the transverse size of the proton while $\Delta$ represents the intrinsic transverse momentum distribution of a parton. In addition, $f_g(x_g)$ and $f_q(x_q)$ are the collinear parton distributions with the parton momentum fractions $x_g$ and $x_q$ for the incoming gluon and quark, respectively. The expression of the $g\to \Q\Qb$ splitting function is
\begin{equation}
k_g^+ \sum_{\alpha\beta\lambda}\psi_{\alpha\beta}^{T\lambda}(\vec r) \psi_{\alpha\beta}^{T\lambda*} (\vec r') = 8\pi^2 m_\Q^2 [(\zeta^2 + (1-\zeta)^2) K_1(m_{\Q} |\vec r|) K_1(m_{\Qb} |\vec r'|) \frac{\vec r\cdot \vec r'}{|\vec r||\vec r'|} + K_0(m_{\Q} |\vec r|) K_0(m_{\Qb} |\vec r'|)]\ .
\end{equation}
Before taking the target expectation value, the scattering amplitudes between the incoming partons (or the final state $ \Q\Qb$ pair) and the background gluon fields in the nuclear target 
$\langle DDD \rangle$ are given by \cite{Zhang:2019dth}
\begin{align}
DDD = [D(\vec x_\Q, \vec x'_\Q) D(\vec x'_{\Qb}, \vec x_{\Qb}) + D(\vec x_g, \vec x'_g) D(\vec x'_g, \vec x_g) - D(\vec x_\Q, \vec x'_g) D(\vec x'_g, \vec x_{\Qb}) - D(\vec x'_{\Qb}, \vec x_g) D(\vec x_g, \vec x'_{\Q})] D(\vec x_q, \vec x'_q),
\end{align}
where the above four terms correspond to four possible diagrams illustrated in Fig.~\ref{fig:feynman}. In the color dipole language, the multiple scatterings between the measured incoming partons and the target can be represented by the product of three dipole amplitudes. In the McLerran-Venugopalan model \cite{McLerran:1993ni,McLerran:1993ka}, the color average expectation of each correlator in $\langle DDD \rangle$ can be evaluated order by order in the large $N_c$ expansion. Up to the $\frac{1}{N_c^2}$ order, the expectation value of the three-dipole amplitude can be written as follows 
\begin{align}
\langle D(\vec x_\Q, \vec x'_\Q) D(\vec x'_{\Qb}, \vec x_{\Qb}) D(\vec x_q, \vec x'_q) \rangle
= 
&
\exp \left[ -\frac{1}{4}Q_s^2 \left((\vec x_\Q - \vec x'_\Q)^2 + (\vec x'_{\Qb} - \vec x_{\Qb})^2 + (\vec x_q - \vec x'_q)^2\right) \right]
\Bigg\{
1 + 
\int_0^1 d\xi \int_0^\xi d\eta  \notag
\\
\Bigg[ 
&
\left(
\frac{Q_s^2}{2N_c} (\vec x_\Q - \vec x'_\Q)\cdot (\vec x_{\Qb} - \vec x'_{\Qb})
\right)^2 
\exp \left(\frac{\eta Q_s^2}{2} (\vec x_\Q - \vec x'_{\Qb}) \cdot (\vec x_{\Qb} - \vec x'_\Q) \right) \notag
\\
+ &
\left(
\frac{Q_s^2}{2N_c} (\vec x_\Q - \vec x'_\Q) \cdot (\vec x_q - \vec x'_q)
\right)^2 
\exp \left (\frac{\eta Q_s^2}{2} (\vec x_\Q - \vec x_q) \cdot (\vec x'_q - \vec x'_\Q) \right) \notag
\\
+ 
&
\left(
\frac{Q_s^2}{2N_c} (\vec x_{\Qb} - \vec x'_{\Qb})\cdot ( \vec x_q - \vec x'_q)
\right)^2 
\exp \left(\frac{\eta Q_s^2}{2} (\vec x'_{\Qb} - \vec x_q) \cdot (\vec x'_q - \vec x_{\Qb})\right)
\Bigg] \Bigg \}, \label{dddq}
\end{align}
where $Q_s$ is the saturation momentum of the large nucleus and the coordinates of the active partons are given by
\begin{align}
 & \vec x_\Q = \vec b_1 + \frac{\vec r_1}{2} + (1-\zeta) \vec r, 
&& \vec x_{\Qb} = \vec b_1 + \frac{\vec r_1}{2} - \zeta \vec r, 
\\
 & \vec x'_{\Q} = \vec b_1 - \frac{\vec r_1}{2} + (1-\zeta) \vec r', 
&& \vec x'_{\Qb} = \vec b_1 - \frac{\vec r_1}{2} -\zeta \vec r', 
\\
 & \vec x_q = \vec b_2 + \frac{\vec r_2}{2},
&& \vec x'_q = \vec b_2 - \frac{\vec r_2}{2},
\\
 & \vec x_g = \vec b_1 + \frac{\vec r_1}{2},
&& \vec x'_g = \vec b_1 - \frac{\vec r_1}{2}.
\end{align}
In order to arrive at the expression in Eq.~(\ref{dddq}), we have neglected the dipole-size logarithmic dependence of the saturation scale, hence obtained the parametrizations valid in a Golec-Biernat-Wusthoff-like approximation. 

Let us also comment on the physical interpretation of each term in Eq.~(\ref{dddq}). The first term, which is the leading $N_c$ contribution, comes from the interaction of three independent dipoles (one for $\Q$, one for $\Qb$ and one for the reference quark) with the dense target gluon fields. This term does not yield any anisotropy since these three dipole amplitudes are uncorrelated in the coordinate space. The other three terms, which are suppressed by $\frac{1}{N_c^2}$, arise when two color singlet dipoles out of those three are broken and converted into a color quadrupole during the interaction with the target nucleus. Due to this peculiar type of color interactions, correlations can be generated between the heavy quark pair and the reference quark in the coordinate space. After the Fourier transform, it is then natural for us to obtain anisotropies for heavy mesons in the momentum space of the order $\frac{1}{N_c^2}$.

\subsection{Differential spectra of heavy quarkonium and open heavy meson together with a reference quark}

Let us first consider the production of the heavy quarkonium. The differential spectrum at the hadronic level is given by the convolution of the partonic spectrum and the corresponding probability density for the hadron of interest. In our study, the probability distribution of producing a heavy-quarkonium from a $\Q\Qb$-pair is given by the color evaporation model (CEM). To generate an open heavy meson, we use the simple Peterson fragmentation function (FF) \cite{Peterson:1982ak} for both $D^0$-meson and $B$-meson, or the KKKS FF for $D^0$ meson \cite{Kneesch:2007ey, Kniehl:2006mw} and KKSS FF for $B$-meson \cite{Kniehl:2008zza}. In the following, we will use the $J/\psi$ and $D^0$ mesons as examples. One can easily get the corresponding formulas for $\Upsilon$ or $B$ meson production by replacing the $c$ quark with a $b$ quark. 

In the CEM, a $c\bar c$-pair can form a $J/\psi$ only if its invariant mass satisfies $M_{J/\psi}^2 < (k_{c} + k_{\bar c})^2 < 4 M_{D}^2$ and the corresponding probability is denoted as $F_{c\bar c\to J/\psi}$. To simplify the multiple dimensional numerical calculation, we take the approximation $\zeta = 1/2$ \cite{Qiu:2013qka}. The kinematic constraint is then given by
\begin{align}
M_{J/\psi}^2-4M_{c}^2 < 4 \Delta \vec k_\perp^2 < 4 M_{D}^2 - 4 M_{c}^2, \label{limits}
\end{align}
where, $\Delta \vec k_\perp = (\vec k_{c\perp}-\vec k_{\bar c\perp})/2$. In our previous study\cite{Zhang:2019dth}, we analytically integrated $\Delta \vec k_\perp$ from $0$ to $\infty$ for the sake of simplicity. We have numerically checked that the above kinematic constraint does not play an important role in the final $v_2$ as shown later in Fig.~\ref{fig:jps-v2}.

The two-particle differential spectrum for the heavy-quarkonium and a reference quark production is
\begin{align}
\left. \frac{dN}{d{\mathcal{PS}}} \right|_{J/\psi+q}
= &
\int d^2k_{c\perp} d^2k_{\bar c\perp} \delta^2 (\vec k_{J/\psi\perp} - \vec k_{c\perp} - \vec k_{\bar c \perp}) \theta (|\Delta \vec k_\perp|-k_{\rm min}) \theta (k_{\rm max} - |\Delta \vec k_\perp|)
\int d\zeta \delta(\zeta-\frac{1}{2}) 
\left. \frac{dN}{d{\mathcal{PS}}} \right|_{c \bar c q} F_{c\bar c\to J/\psi}
\nonumber \\
= &
\mathcal{N}_{J/\psi}
\int \frac{d^2b_1 d^2b_2 d^2r_1 d^2r_2 d^2rd^2r^\prime}{(2\pi)^8} \int_{k_{\rm min}}^{k_{\rm max}} d\Delta k_{\perp} \Delta k_\perp J_0(\Delta k_\perp |\vec r - \vec r'|) e^{-i \vec k_{J/\psi \perp} \cdot \vec r_1} e^{-i \vec k_{q\perp} \cdot \vec r_2}
\nonumber \\
&
\times
W(x_g,x_q, b_1,b_2,r_1,r_2) 
k_g^+ \sum_{\alpha\beta\lambda}
\left.
\psi_{\alpha\beta}^{T\lambda}(\vec r) \psi_{\alpha\beta}^{T\lambda*} (\vec r')
\right|_{\zeta=\frac{1}{2}}
\left.
\langle DDD \rangle
\right|_{\zeta=\frac{1}{2}} ,
\label{eq:cs-jpsi}
\end{align}
where $d{\mathcal{PS}} = dy_{J/\psi} d^2k_{J/\psi\perp} dy_q d^2k_{q\perp}$ is the final state phase space for this process, $k_{\rm min} = 0.98$ GeV and $k_{\rm max} = 2.95$ GeV are computed from Eq.~(\ref{limits}). In our calculation, we take $F_{c\bar c\to J/\psi}$ as a constant which can be absorbed into the new overall normalization factor $\mathcal{N}_{J/\psi}$. The overall normalization will disappear in the calculation of the elliptic flow due to cancellation in ratios. The full phase space integration of $d^2\Delta k_\perp$ yields a delta function, $\delta^2 (\vec r-\vec r')$, which allows us to get rid of the $d^2r'$ integral. In this paper, we carry out a more sophisticated calculation by using Eq. (\ref{eq:cs-jpsi}).

As to the case of the open heavy flavor meson, such as the $D$ meson, the differential spectrum for $D^0+q$ production process can be obtained from Eq.~(\ref{eq:cs-qqq}) by integrating over the phase space of $\bar c$ quark and convoluting with the $c\to D^0$ fragmentation function, $D(z)$. The full space integral of $d^2k_{\bar c\perp}$ generates a delta function, $(2\pi)^2\delta^2 (\vec r_1- \zeta (\vec r - \vec r'))$, which removes the $d^2r'$ integral. The final expression is then given by
\begin{align}
\left. \frac{dN}{d{\mathcal{PS}}} \right|_{D^0+q}
=
&
\int d\zeta d^2k_{\bar c\perp} \int dz \frac{D(z)}{z^2} 
\left. \frac{dN}{d{\mathcal{PS}}} \right|_{c \bar c q}
\nonumber \\
=
&
\mathcal{N}
\int d\zeta \int dz \frac{D(z)}{z^2} 
\int \frac{d^2b_1 d^2b_2 d^2r_1 d^2r_2 d^2r}{(2\pi)^7} 
\frac{1}{\zeta^2}
e^{-i \frac{\vec k_{D\perp}}{z} \cdot \frac{\vec r_1}{\zeta}} e^{-i\vec k_{q\perp} \cdot \vec r_2} W(x_g,x_q,b_1,b_2,r_1,r_2) 
\nonumber \\
&
\times
k_g^+ \sum_{\alpha\beta\gamma}\psi_{\alpha\beta}^{T\lambda}(\vec r) \psi_{\alpha\beta}^{T\lambda*} (\vec r'=\vec r-\frac{\vec r_1}{\zeta}) \langle DDD \rangle |_{\vec r'=\vec r-\frac{\vec r_1}{\zeta}},
\label{eq:cs-d0}
\end{align}
where the phase space for this process is $d{\mathcal{PS}} = dy_{D}d^2k_{D\perp} dy_q d^2k_{q\perp}$ and $\vec k_{c\perp} = \vec k_{D\perp}/z$.

\subsection{Elliptic flow of heavy-quarkonium and open heavy-meson}

Following the experimental setup and the usual convention, we define the transverse momentum dependent $n$-th Fourier harmonic from the differential spectrum of particle $X$ plus the reference quark production as \cite{Borghini:2001vi}
\begin{align}
\frac{d\kappa_n}{dy_X dk_{X\perp}} = k_{X\perp} \int d\phi_X dy_q d^2k_{q\perp} e^{-in(\phi_X-\phi_q)} 
\left. \frac{dN}{d{\mathcal{PS}}}\right|_{X+q}.
\label{eq:kappan}
\end{align}
Using the commonly used two-particle correlation method adopted by the CMS collaboration\cite{CMS:2018xac} for heavy meson flows, the elliptic flow can be computed as follows 
\begin{align}
v_2 (y_X, k_{X\perp}) = \frac{d\kappa_2/dy_X dk_{X\perp}}{d\kappa_0/dy_X dk_{X\perp}} \frac{1}{v_2 [{\rm ref}]},
\label{eq:v2-def}
\end{align}
where $v_2[{\rm ref}]$ is the integrated elliptic flow of the reference quark given by $v_2[{\rm ref}] = \sqrt{\kappa_2[{\rm ref}]/\kappa_0[{\rm ref}]}$. In the following, we will present the expression for the second and zeroth harmonics for the cases of $J/\psi+q$ and $D_0+q$ productions. Then, the corresponding elliptic flow can be easily obtained from Eq. (\ref{eq:v2-def}).

Substituting Eq. (\ref{eq:cs-jpsi}) into Eq. (\ref{eq:kappan}), we can obtain the second and zeroth harmonics that are required to calculate the elliptic flow of heavy quarkonium. Since there are many dimensions of integrations, in order to obtain elliptic flows numerically, our strategy is to analytically perform as many integrations as possible and evaluate the rest of the integrations numerically. The second harmonic is given by
\begin{align}
\frac{d\kappa_2}{dy_{J/\psi}dk_{J/\psi \perp}}
=  &
k_{J/\psi \perp} \mathcal{N}_{J/\psi} \int dy_q k_{q\perp} dk_{q\perp} \int \frac{r_1 dr_1 d^2r_2 d^2r d^2r'}{(2\pi)^3} \int_0^1  d\xi \int_0^\xi d\eta J_2 (|\vec k_{J/\psi\perp}||\vec r_1|) J_2 (|\vec k_{q\perp}||\vec r_2|)
\cos (2\phi_{r_2}) 
\nonumber \\
& \times 
\frac{k_{\rm max} J_1(k_{\rm max}|\vec r - \vec r'|) - k_{\rm min} J_1(k_{\rm min}|\vec r - \vec r'|)}{|\vec r - \vec r'|}
x_g f_g(x_g) x_q f_q(x_q) \frac{1}{4\pi^2} \exp\left[-\frac{\Delta^2 (r_1^2+r_2^2)}{4}\right] 
\frac{1}{1+\eta Q_s^2 B_p}
\nonumber \\
& \times
k_g^+ \sum_{\alpha\beta\gamma} \left. \psi_{\alpha\beta}^{T\lambda} (\vec r) \psi_{\alpha\beta}^{T\lambda*} (\vec r')\right|_{\zeta = \frac{1}{2}}
\frac{Q_s^4}{4N_c^2} \Bigl[ 
\mathcal{F}_1 (r_1, r_2, r, r') + \mathcal{F}_2 (r_1, r_2) - \mathcal{F}_3 (r_1, r_2, r) - \mathcal{F}_4 (r_1, r_2, r')
\Bigr],
\label{eq:kappa2-jpsi}
\end{align}
where $J_{i}$'s are Bessel functions of the first kind and $\mathcal{F}_{i}$'s are defined as
\begin{align}
& \mathcal{F}_1 (r_1, r_2, r, r') 
=  
2 \left[ (\vec r_1 + \frac{\vec r - \vec r'}{2}) \cdot \vec r_2\right]^2 
\exp \left[ - \frac{\eta Q_s^2 (\vec r + \vec r')^2}{32 (1+ \eta Q_s^2 B_p)} \right] 
\exp \left[ - \frac{Q_s^2}{4} (2r_1^2 + r_2^2 + \frac{(\vec r- \vec r')^2}{2}) \right]
\nonumber \\
&
\phantom{XXXXXXX}
\times 
\exp \left[ \frac{\eta Q_s^2 (\vec r_1 - \vec r_2 + \frac{\vec r - \vec r'}{2})^2}{8} \right],
\\
& \mathcal{F}_2 (r_1, r_2) 
= 
2 (\vec r_1 \cdot \vec r_2)^2 
\exp \left[ - \frac{Q_s^2}{4} (2r_1^2 + r_2^2) \right] 
\exp \left[ \frac{\eta Q_s^2}{8} (\vec r_1 - \vec r_2)^2 \right],
\\
& \mathcal{F}_3 (r_1, r_2, r)
= 
2 \left( (\vec r_1 + \frac{\vec r}{2}) \cdot \vec r_2 \right)^2
\exp \left[ - \frac{\eta Q_s^2 r^2}{32 (1+\eta Q_s^2 B_p)} \right]
\exp \left[ - \frac{Q_s^2}{4} (2r_1^2 + r_2^2 + \frac{r^2}{2}) \right]
\nonumber \\
& 
\phantom{XXXXXXX}
\times 
\exp \left[ \frac{\eta Q_s^2}{8} (\vec r_1 - \vec r_2 + \frac{\vec r}{2})^2 \right],
\\
& \mathcal{F}_4 (r_1, r_2, r') = \mathcal{F}_3 (r_1,r_2, r').
\end{align}
Since the $dy_q$ and $d^2k_{q\perp}$ integrals approximately factorize, one writes $\int dy_q x_q f_q (x_q) = \int dx_q f(x_q)$ which is just total number of the reference quark. To obtain the corresponding $v_2$, we have to rely on the numerical evaluation of the rest of the 10-dimension integral. 

The zeroth Fourier harmonic on the other hand is relatively simple. The $d^2r_2$ integral is eliminated by the delta function obtained from the $d^2k_{q\perp}$ integration. The integrals over $d^2b_1d^2b_2d^2r_1$ can be carried out analytically. For self-consistency in terms of the large $N_c$ expansion, we only need to keep the leading-$N_c$ contributions as well. In the end, we only need to numerically compute a three-dimentional integral, which is given by
\begin{align}
\frac{d\kappa_0}{dy_{J/\psi}dk_{J/\psi \perp}}
= ~ & 
k_{J/\psi \perp} \mathcal{N}_{J/\psi} \int dy_q \frac{r dr d^2 r'}{2\pi} 
\frac{k_{\rm max} J_1(k_{\rm max}|\vec r - \vec r'|) - k_{\rm min} J_1(k_{\rm min}|\vec r - \vec r'|)}{|\vec r - \vec r'|} 
\frac{1}{4\pi^2} 
\frac{2}{\Delta^2 + 2Q_s^2}
\nonumber \\
& \times
x_g f_g(x_g) x_q f_q(x_q) 
k_g^+ \sum_{\alpha\beta\gamma} \left. \psi_{\alpha\beta}^{T\lambda} (\vec r) \psi_{\alpha\beta}^{T\lambda*} (\vec r')\right|_{\zeta = \frac{1}{2}} \exp[- \frac{k_{J/\psi\perp}^2}{\Delta^2 + 2 Q_s^2}]
\nonumber \\
& \times \left\{ \exp \left[-\frac{Q_s^2(\vec r - \vec r')^2}{8}\right] + 1 - \exp \left[-\frac{Q_s^2}{8} r^2\right] - \exp \left[-\frac{Q_s^2}{8} r'^{2}\right] \right\}.
\label{eq:kappa0-jpsi}
\end{align}
It is easy to note that the integral $\int dy_q x_q f_q(x_q)$ gives an overall factor which cancels the identical one in $\kappa_2$, together with the overall constant.

For the open heavy meson production accompanied by a reference quark process, we can simplify the expression by using several tricks which are provided in Appendix \ref{ch:kappa2-d0}. These procedures yield the expression which is less time consuming and more accurate in the numerical evaluation. The second harmonic of the differential spectrum of the open heavy meson plus a reference quark production is given by
\begin{align}
\frac{d\kappa_2}{dy_{D}dk_{D \perp}}
= ~ &
k_{D\perp} \mathcal{N} \int dy_q k_{q\perp} dk_{q\perp}\int d\zeta \int \frac{r_1 dr_1 d^2 r}{2\pi}
\frac{1}{\zeta^2} \int dz \frac{D(z)}{z^2} \int d\xi \int_0^\xi d\eta J_2 (\frac{|\vec k_{D\perp}||\vec r_1|}{z \zeta}) 
\nonumber \\
& \times
x_g f_g(x_g) x_q f_q(x_q) \frac{1}{4\pi^2} e^{-\frac{\Delta^2 r_1^2}{4}}
\frac{1}{1+\eta Q_s^2 B_p} 
k_g^+ \sum_{\alpha\beta\lambda} \left.\psi_{\alpha\beta}^{T\lambda}(\vec r) \psi_{\alpha\beta}^{T\lambda*} (\vec r') \right|_{\vec r'=\vec r-\frac{\vec r_1}{\zeta}}
\nonumber \\ 
& \times 
\frac{Q_s^4}{4N_c^2} 
[\mathcal{F}^{D}_1 (r_1,r) + \mathcal{F}^{D}_2 (r_1) - \mathcal{F}^{D}_3 (r_1,r) - \mathcal{F}^{D}_4 (r_1,r)],
\label{eq:kappa2-d0}
\end{align}
where the $\mathcal{F}^{D}_i$'s are defined as
\begin{align}
& \mathcal{F}^{D}_1 (r_1, r) = 
\exp\left[ - \frac{\eta Q_s^2}{2(1+ \eta Q_s^2 B_p)} ((1-\zeta)\vec r - \frac{1-\zeta}{2\zeta} \vec r_1)^2 \right]
\exp \left[ -\frac{(2-\eta) Q_s^2 r_1^2}{8\zeta^2} \right]
\nonumber\\
&
\phantom{XXXXXX} 
\times
\sum_{m=0}^\infty \left[ \frac{\eta Q_s^2}{4}\right]^{2m}
\left(\frac{r_1}{\zeta}\right)^{2m+2} \frac{(2m+2)! k_{q\perp}^2}{2^{2m+6} a_q^{m+3} (2m)! m!}
{}_1F_1 (m+3, 3, -\frac{k_{q\perp}^2}{4a_q}),
\label{eq:fd-1}
\\
& \mathcal{F}^{D}_2 (r_1) =
\exp \left[ - \frac{(4-\eta) Q_s^2}{8} r_1^2 \right]
\sum_{m=0}^\infty \left[\frac{\eta Q_s^2}{4}\right]^{2m}
2 r_1^{2m+2} 
\frac{(2m+2)! k_{q\perp}^2}{2^{2m+6} a_q^{m+3} (2m)! m!} 
{}_1F_1 (m+3, 3, -\frac{k_{q\perp}^2}{4a_q}),
\label{eq:fd-2}
\\
& \mathcal{F}^{D}_3 (r_1, r) = 
\exp \left[ -\frac{Q_s^2}{4} (2r_1^2 + (\zeta^2+(1-\zeta)^2) r^2 + 2 (1-2\zeta) \vec r \cdot \vec r_1) \right]
\exp \left[ -\frac{\eta Q_s^2}{8(1+\eta Q_s^2 B_p)} (1-\zeta)^2 r^2 \right]
\nonumber \\
& 
\phantom{XXXXXX} \times
\exp \left[  \frac{\eta Q_s^2}{8} (\vec r_1 +(1-\zeta)\vec r)^2 \right]
\cos(2\phi_A)
\nonumber\\
&
\phantom{XXXXXX} \times
\sum_{m=0}^\infty \left[ \frac{\eta Q_s^2}{4} \right]^{2m} |\vec r_1 + (1-\zeta) \vec r |^{2m+2} 
\frac{(2m+2)! k_{q\perp}^2}{2^{2m+6} a_q^{m+3} (2m)! m!}
{}_1F_1 (m+3, 3, -\frac{k_{q\perp}^2}{4a_q})
\nonumber\\
&
\phantom{XXXXXX} 
+ 
\exp \left[ -\frac{Q_s^2}{4} (2r_1^2 + (\zeta^2+(1-\zeta)^2) r^2 + 2(1-2\zeta) \vec r \cdot \vec r_1) \right]
\exp \left[ -\frac{\eta Q_s^2}{8(1+\eta Q_s^2 B_p)} \zeta^2 r^2  \right]
\nonumber\\
&
\phantom{XXXXXX} 
\times
\exp \left[ \frac{\eta Q_s^2}{8} (\vec r_1 - \zeta \vec r)^2 \right] 
\cos(2\phi_B) 
\nonumber \\
&
\phantom{XXXXXX} 
\times
\sum_{m=0}^\infty \left[ \frac{\eta Q_s^2}{4} \right]^{2m} |\vec r_1 - \zeta \vec r|^{2m+2}
\frac{(2m+2)! k_{q\perp}^2}{2^{2m+6} a_q^{m+3} (2m)! m!} 
{}_1F_1 (m+3, 3, -\frac{k_{q\perp}^2}{4a_q}),
\label{eq:fd-3}
\\
&
\mathcal{F}^{D}_4 (r_1,r)
=
\exp \Bigl[-\frac{Q_s^2}{4} [(\frac{\vec r_1}{\zeta}-(1-\zeta) \vec r)^2 + \zeta^2 r^2] \Bigr]
\exp \left[-\frac{\eta Q_s^2}{8(1+\eta Q_s^2 B_p)} (\vec r_1-\zeta \vec r)^2 \right]
\exp \left[\frac{\eta Q_s^2}{8} \zeta^2 r^2 \right]
\cos(2\phi_r)
\nonumber \\
& 
\phantom{XXXXXX}
\times
\sum_{m=0}^\infty \left[ \frac{\eta Q_s^2}{4} \right]^{2m} 
|\zeta r|^{2m+2} 
\frac{(2m+2)! k_{q\perp}^2}{2^{2m+6} a_q^{m+3} (2m)! m!}
{}_1F_1 (m+3, 3, -\frac{k_s^2}{4a_q})
\nonumber \\
&
\phantom{XXXXXX}
+
\exp \Bigl[-\frac{Q_s^2}{4} [(\frac{\vec r_1}{\zeta}-(1-\zeta) \vec r)^2 + \zeta^2 r^2] \Bigr]
\exp \left[-\frac{\eta Q_s^2 (1-\zeta)^2}{8(1+\eta Q_s^2 B_p)} (\vec r-\frac{\vec r_1}{\zeta})^2 \right]
\nonumber \\
&
\phantom{XXXXXX}
\times
\exp \left[\frac{\eta Q_s^2}{8} (\frac{\vec r_1}{\zeta} -(1-\zeta) \vec r)^2 \right] 
\cos(2\phi_{C})
\nonumber \\
& 
\phantom{XXXXXX}
\times
\sum_{m=0}^\infty 
\left[ \frac{\eta Q_s^2}{4} \right]^{2m}
|\frac{\vec r_1}{\zeta}  - (1-\zeta) \vec r|^{2m+2}
\frac{(2m+2)! k_{q\perp}^2}{2^{2m+6} a_q^{m+3} (2m)! m!}
{}_1F_1 (m+3, 3, -\frac{k_{q\perp}^2}{4a_q}),
\label{eq:fd-4}
\end{align}
with $a_q = [2\Delta^2+(2-\eta)Q_s^2]/8$, $\phi_A$ the angle between $\vec r_1$ and $\vec r_1 + (1-\zeta) \vec r$, $\phi_B$ the angle between $\vec r_1$ and $\vec r_1 - \zeta \vec r$ and $\phi_C$ the angle between $\vec r_1$ and $\vec r_1/\zeta - (1-\zeta) \vec r$. In fact, $\mathcal{F}^{D}_3 (r_1,r)$ and $\mathcal{F}^{D}_4 (r_1,r)$ give exactly the same contribution. This symmetry also indicates that the elliptic flow of the $\bar c$ quark or $\bar D^0$ meson are exactly the same with that of the $c$ quark or $D^0$ meson.

Employing the same tricks presented in Appendix \ref{ch:kappa2-d0}, we can carry out the $d^2b_1 d^2b_2$ integration analytically for the zeroth harmonic contribution. It is then given by
\begin{align}
\frac{d\kappa_0}{dy_{D}dk_{D \perp}}
= ~ &
k_{D\perp} \mathcal{N} 
\int dy_q \int \frac{r_1dr_1d^2r}{2\pi} \int \frac{d\zeta}{\zeta^2} \int dz \frac{D(z)}{z^2}  J_0 (\frac{|\vec k_{D\perp}||\vec r_1|}{z \zeta}) 
\nonumber\\
& \times
x_g f_g(x_g) x_q f_q(x_q) \frac{1}{4\pi^2} e^{-\frac{\Delta^2 r_1^2}{4}}
k_g^+ \sum_{\alpha\beta\lambda} \left.\psi_{\alpha\beta}^{T\lambda}(\vec r) \psi_{\alpha\beta}^{T\lambda*} (\vec r') \right|_{\vec r'=\vec r-\frac{\vec r_1}{\zeta}}
\nonumber\\
& \times
\left\{\exp\left[-\frac{Q_s^2 r_1^2}{4\zeta^2}\right]
+ \exp\left[-\frac{Q_s^2 r_1^2}{2}\right]
- \exp\left[ - \frac{Q_s^2}{4} (2r_1^2 + (\zeta^2+(1-\zeta)^2)r^2 + 2 (1-2\zeta) \vec r_1 \cdot \vec r) \right] \right.
\nonumber \\
& \phantom{XXX}
\left. - \exp\left[ - \frac{Q_s^2}{4} (\frac{r_1^2}{\zeta^2} + (\zeta^2+(1-\zeta)^2) r^2 - \frac{2(1-\zeta)}{\zeta} \vec r_1 \cdot \vec r) \right] \right\}.
\label{eq:kappa0-d0}
\end{align}
The above expressions allow us to compute the second harmonics for both heavy quarkonia and open heavy meson similar to the measurement conducted at the LHC. 

\subsection{Matching to the cross section of single inclusive particle production} 

In addition, if one integrates over the phase space of the reference quark, it is natural to find that the zeroth harmonic of two-particle spectrum becomes the differential spectrum of single inclusive particle production. Therefore, we should expect that $d\kappa_0/dydk_\perp$ matches to the single inclusive particle cross section, if the overall normalization factor is properly recovered.

The differential cross section for single inclusive $J/\psi$ production can be found in Refs.~\cite{Ma:2014mri, Ma:2015sia, Watanabe:2015yca}. As is pointed out in those works, the CGC framework can only describe the low transverse momentum spectrum of the heavy quarkonium production when $p_T\leq Q_s$, while the high transverse momentum region is described by higher-order QCD calculations in the collinear framework.

In contrast, the situation of the open heavy meson is a bit different: the other heavy quark is unobserved and it can provide sufficient amount of transverse momentum recoils without the need of additional gluon radiation. Usually a simple leading order CGC model calculation with properly chosen saturation momentum can describe the spectrum of open heavy quark or meson from low $p_T$ to high $p_T$ region. We find that the inclusive $D^0$ meson cross section in the dilute-dense factorization framework is \cite{Blaizot:2004wv,Fujii:2006ab,Dominguez:2011wm,Fujii:2013yja,Ma:2018bax}
\begin{align}
\frac{d\sigma}{dy_{D} d^2k_{D\perp}} = 
~& \alpha_s T_R \frac{S_\perp}{(2\pi)^2} \int \frac{d^2r_1 d^2r}{(2\pi)^4} \int dz \frac{D(z)}{z^2} \int \frac{d\zeta}{\zeta^2} e^{-i \vec k_{D\perp}\cdot \vec r_1 /z\zeta } x_g f_g(x_g) 
k_g^+ \sum_{\alpha\beta\lambda} \left.\psi_{\alpha\beta}^{T\lambda}(\vec r) \psi_{\alpha\beta}^{T\lambda*} (\vec r') \right|_{\vec r'=\vec r-\frac{\vec r_1}{\zeta}}
\nonumber\\
& \times
\left\{
\exp\left[-\frac{Q_s^2 r_1^2}{4\zeta^2}\right]
+ \exp\left[-\frac{Q_s^2 r_1^2}{2}\right]
- \exp\left[ - \frac{Q_s^2}{4} (2r_1^2 + (\zeta^2+(1-\zeta)^2)r^2 + 2 (1-2\zeta) \vec r_1 \cdot \vec r) \right]  \right.
\nonumber \\
& \phantom{XXX}
\left.- \exp\left[ - \frac{Q_s^2}{4} (\frac{r_1^2}{\zeta^2} + (\zeta^2+(1-\zeta)^2) r^2 - \frac{2(1-\zeta)}{\zeta} \vec r_1 \cdot \vec r) \right]
\right\}.
\label{eq:dilute-dense-D}
\end{align}
where $S_\perp$ is the effective area of the target hadron. If we set $\mathcal{N} = \frac{\alpha_s T_R S_\perp}{2\pi}$ and remove the reference quark number $\int dy_q x_q f_q(x_q)$, Eq. (\ref{eq:kappa0-d0}) and Eq. (\ref{eq:dilute-dense-D}) only differ from each other in the Gaussian distribution $e^{-\frac{\Delta^2 r_1^2}{4}}$, which is numerically insignificant. 

As shown in Fig.~\ref{fig:cb-spectra}, even with the Gaussian parameterization of the scattering amplitude, we can obtain a good description of the $B$ meson spectrum measured by CDF by using Eq. (\ref{eq:dilute-dense-D}). This may imply that we could push our $v_2$ results for open heavy flavor to a regime of higher $p_T$.

\end{widetext}

\section{Numerical results}

In this section, we provide the results of the numerical evaluation of the heavy meson $v_2$ derived in previous sections. Together with the input of parton distribution functions for the incoming proton\cite{Martin:2009iq}, we can describe the collective behavior of heavy quarkonia and open heavy flavors with the same CGC model, which indicates the robustness of the anisotropy generated from the initial state effects in the CGC formalism in small collision systems. 

\subsection{Elliptic flow of heavy quarkonia}

In Ref.~\cite{Zhang:2019dth}, we have presented the numerical results of elliptic flow of heavy quarkonia by analytically integrating over the relative momentum between the heavy quark pair $\Q\Qb$ from $0$ to $\infty$. This is an approximation which does not affect the resulting $v_2$ as we show in the following numerical calculations.  

\begin{figure}[!h]
\includegraphics[width=0.45\textwidth]{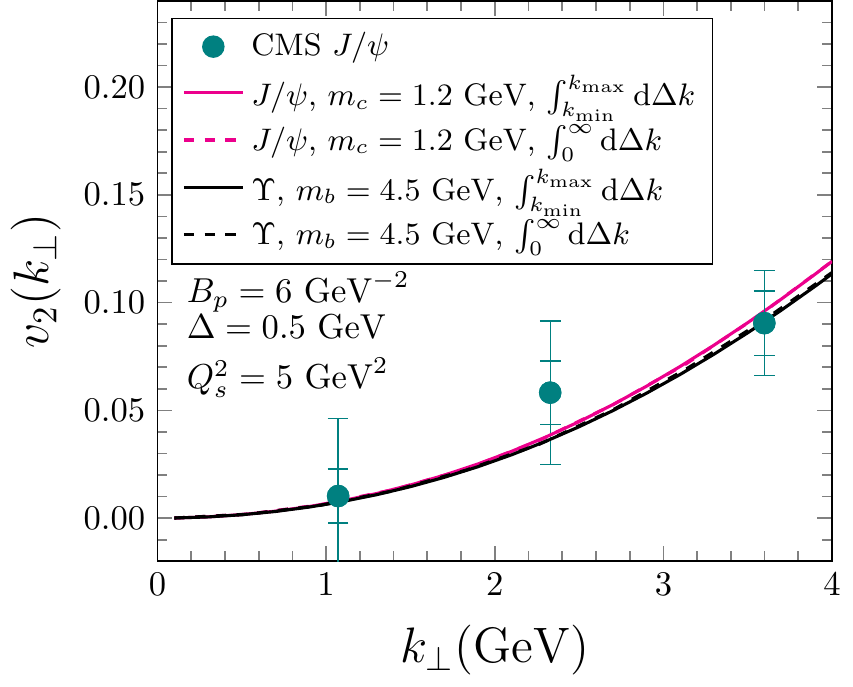}
\caption{Elliptic flow of heavy quarkonium in $g\to\Q\Qb +q$ channel compared with the experimental data from the CMS collaboration \cite{CMS:2018xac}.}
\label{fig:jps-v2}
\end{figure}

In this work, we perform a more sophisticated calculation with the proper kinematic constraints implied in the CEM using Eqs. (\ref{eq:kappa2-jpsi}) and (\ref{eq:kappa0-jpsi}). As suggested in Ref. \cite{Zhang:2019dth}, we indeed find little difference between the two calculations as shown in Fig. \ref{fig:jps-v2}. This indicates that the kinematic constraints, which reduce the yield of heavy quarkonia, have little impact on the angular correlation such as $v_2$. Similarly to the results in Ref. \cite{Zhang:2019dth}, a very weak mass dependence of the heavy quarkonium $v_2$ is found in the numerical evaluation, mainly due to the fact that the mass dependent terms in $\kappa_2$ and $\kappa_0$ cancel each other in the leading small $r^2$ expansions where $r$ is the coordinate separation of the $\Q\Qb$ pair.

\begin{figure}[!h]
\includegraphics[width=0.45\textwidth]{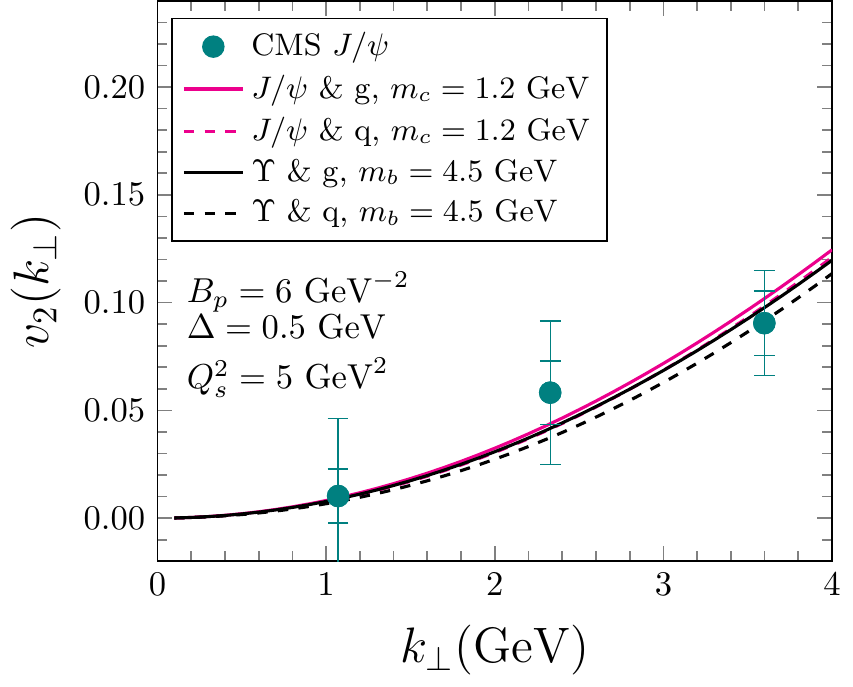}
\caption{Elliptic flows of heavy quarkonia with a quark or a gluon as the reference particle.}
\label{fig:jpsi-qg-and-gg}
\end{figure}

To compare with the experimental data which measures the correlation between $J/\Psi$ and a charged hadron which serves as the reference particle, we need to take into account both $g \to\Q\Qb +q$ and $g\to \Q\Qb+g$ channels since the charged hadron can be fragmented either from a quark or a gluon. When the reference quark is replaced by a reference gluon, we need to compute a slightly different set of diagrams. The elliptic flow of heavy quarkonia from the $g\to \Q\Qb+g$ channel has also been studied and the numerical results are shown in Fig. \ref{fig:jpsi-qg-and-gg}. Since the difference between these two channels are negligible, we do not expect a significant modification on the final results. In addition, we have also found that the use of the charged-hadron fragmentation function \cite{Albino:2008fy} does not affect the final results for heavy quarkonium flow, since we also need to integrate over the phase space of the charged-hadron as indicated in the measurement.

As pointed out in Ref. \cite{Zhang:2019dth}, our above calculation for heavy quarkonia is valid only in the low transverse momentum region for several reasons. At leading order in the CGC framework, the transverse momentum of final state heavy quarkonium receives transverse momentum contribution only from the transverse momenta of the incoming partons. In the large $p_T$ region, where a turnover in $v_2$ as a function of the transverse momentum should occur, our simple parametrization of the Golec-Biernat and Wusthoff type Gaussian distribution\cite{GolecBiernat:1998js} becomes insufficient to describe the heavy quarkonia $p_T$ spectrum. Instead, we would need to employ a more accurate and sophisticated implementation of small-x dipole amplitudes, such as the numerical solution to the small-$x$ evolution equations. Furthermore, as shown in Ref.~\cite{Ma:2014mri, Ma:2015sia}, one needs to take the extra gluon radiation into account in order to generate large momentum recoils in the high $p_T$ region. Interestingly, the situation changes for the open heavy meson production as we discuss below.

\subsection{Elliptic flow of open heavy mesons}

\begin{figure}[h!]
\includegraphics[width=0.45\textwidth]{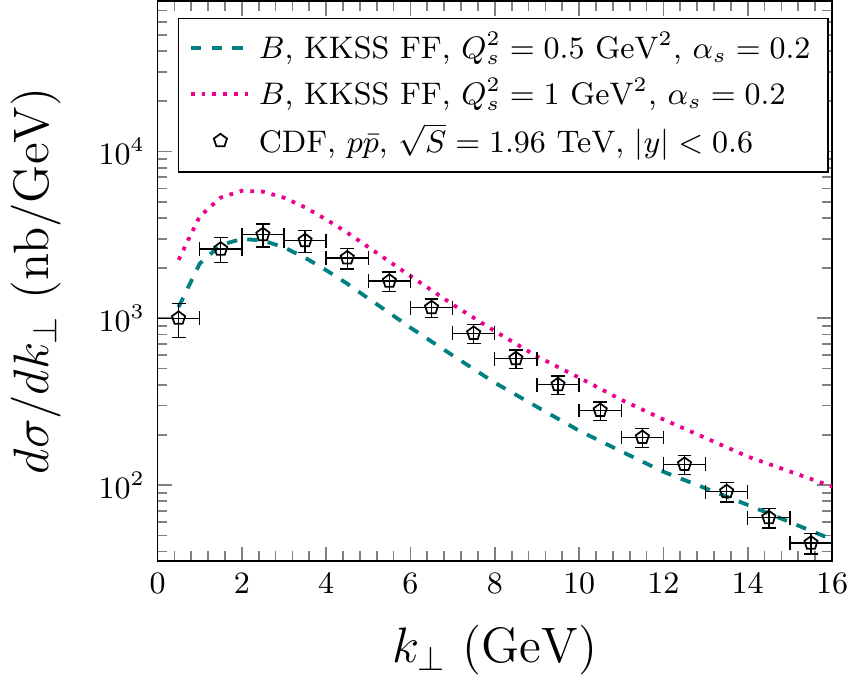}
\caption{Differential cross section for $B$ mesons computed in our model and compared with experimental data from the CDF collaboration \cite{Acosta:2004yw} in $p\bar p$ collisions. Note that we use smaller values of $Q_s^2$ due to the consideration of smaller target size and lower collision energy and $S_\perp = 12$ mb for a proton target.}
\label{fig:cb-spectra}
\end{figure}

For the case of the open heavy flavor meson production, we only measure one heavy quark and integrate over the phase space of the other one. The transverse momentum distribution of the final state open heavy meson at large transverse momentum is mostly controlled by the hard $g\to \Q\Qb$ splitting, which can provide sufficiently large momentum recoils. Although the total transverse momentum of the $ \Q\Qb$ pair is predominantly small, each individual heavy quark can have a large transverse momentum due to the hard $g\to \Q\Qb$ splitting. The situation is similar to the inclusive hadron or jet production in the collinear factorization framework, in which the leading order calculation can already provide a good description of the transverse momentum distribution. For example, the transverse momentum spectrum of B mesons in $p\bar p$ collision can be computed and the results are shown in Fig. \ref{fig:cb-spectra}. Within our simplified CGC model, the shape of the experimental data reported in Ref.~\cite{Acosta:2004yw} in both low and high $p_T$ regime can be described with a corresponding saturation momentum for the proton target.\footnote{Strictly speaking, our model is more applicable to $pA$ collisions. Nevertheless, the comparison shown in Fig.~\ref{fig:cb-spectra} serves as a quantitative example in order to demonstrate our discussion regarding the high transverse momentum region.} We expect that a better agreement could be reached if one uses the numerical solution to the small-$x$ evolution equation instead of the Golec-Biernat and Wusthoff type Gaussian distribution. This implies that we may be able to extend the region of validity of our calculation to high transverse momentum for the open heavy flavor meson, as we show below. 

\begin{figure}[!h]
\includegraphics[width=0.45\textwidth]{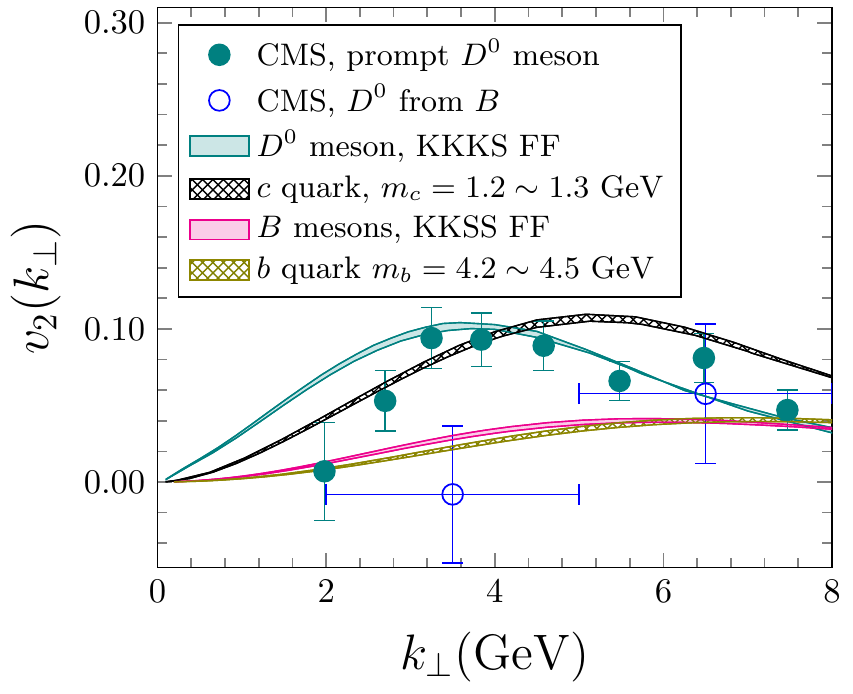}
\caption{Elliptic flow of open heavy mesons in the $g\to \Q\Qb+q$ channel compared with the experimental data from the CMS collaboration \cite{Sirunyan:2018toe, CMS:2019isc}. In the numerical calculation, the same set of values for parameters have been adopted than in the quarkonia case, namely, $B_p = 6$ GeV$^{-2}$, $\Delta = 0.5$ GeV and $Q_s^2 = 5$ GeV$^2$.}
\label{fig:v2-d0}
\end{figure}

\begin{figure}[!h]
\includegraphics[width=0.45\textwidth]{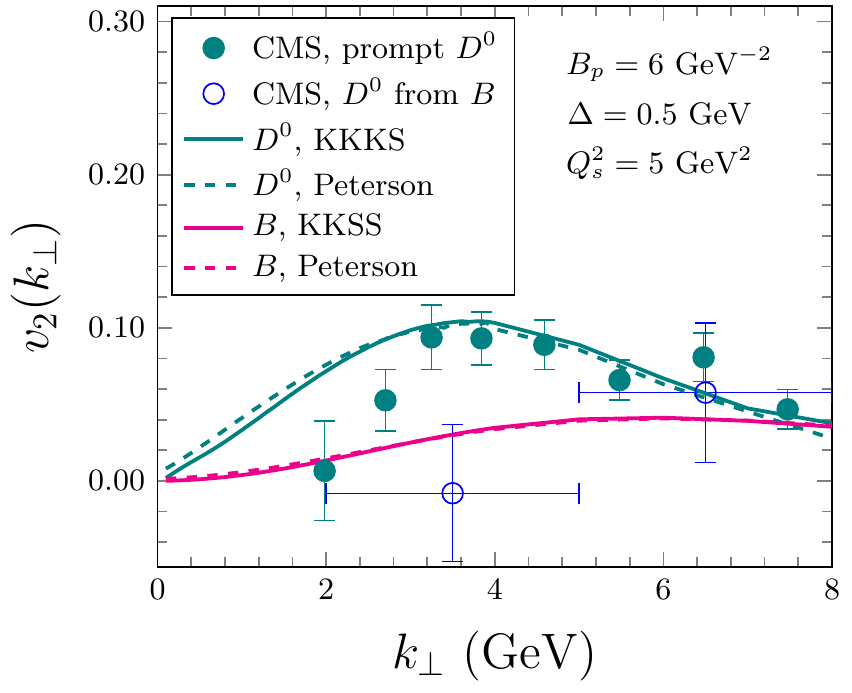}
\caption{Elliptic flows of open heavy mesons in the $g\to \Q\Qb+q$ channel calculated with different fragmentation functions.}
\label{fig:v2-d0-ffs}
\end{figure}

Using the same set of parameters as in the calculation for heavy quarkonia, we numerically compute the $v_2$ of open heavy mesons up to $8$ GeV in $g+q$ channel using Eqs. (\ref{eq:kappa2-d0}) and (\ref{eq:kappa0-d0}) and show the results in Figs. \ref{fig:v2-d0} and \ref{fig:v2-d0-ffs}. In our numerical evaluation, we have adopted the FFs provided by the Peterson model \cite{Peterson:1982ak} for both $D^0$-meson and $B$-meson, and also the KKKS FF for $D^0$ meson \cite{Kneesch:2007ey, Kniehl:2006mw} and the KKSS FF for $B$-meson \cite{Kniehl:2008zza}. There is a clear shift from the $v_2$ of $c$ quark to that of the $D^0$ meson while it is less obvious for the $v_2$ of the $b$ quark and the $B$ meson. This is mainly due to the $b\to B$ fragmentation function, which is strongly peaked at a larger value of $z$ compared to the $c\to D^0$ one. As shown in Fig. \ref{fig:v2-d0-ffs}, we find that the resulting $v_2$ is insensitive to the choices of the FFs.

As shown in Fig.~\ref{fig:v2-d0}, our calculation of the $v_2$ for $D^0$ meson production can describe the CMS data \cite{Sirunyan:2018toe} reasonably well within the uncertainties of the experimental data. In addition, we can make prediction for the second harmonic coefficients of $b$ quarks and $B$ mesons as well, which are strongly suppressed as compared to those coefficients of $D$ mesons and heavy quarkonia. In contrast to the heavy quarkonia case, we observed a strong mass dependence for the open heavy meson $v_2$ in our theoretical and numerical calculations. For the $v_2$ of heavy quarkonia, the mass dependent term coming from the splitting function is only present in the $d^2r$ and $d^2r'$ integrals which can be factorized out from the integration of azimuthal angle of the reference quark. Therefore, it contributes little to the elliptic flow. Physically speaking, since we always require the heavy $\Q\Qb$ pair to be close together in order to produce the quarkonium, the splitting process does not really modify the correlation between the $\Q\Qb$ pair and the reference quark. For the Fourier harmonics of open heavy mesons, as can be seen from Eqs. (\ref{eq:fd-3}-\ref{eq:fd-4}), the $d^2r$, $d^2r'$ and $d^2r_2$ integrals are entangled together. In this case, we are studying the correlation between one final state heavy quark out of the splitting and a reference parton. Since the distance between the $\Q\Qb$ pair can be arbitrarily large, the mass dependence naturally comes in the correlation. In addition, we know that usually the mass of heavy quarks always contributes as a suppression in the propagator, and we also note the scale ordering $m_c <Q_s < m_b$. Therefore, it is reasonable to expect that the $D$ meson can have a sizable $v_2$ coefficient, while the correlation between the $B$ meson and the reference particle should be suppressed. 

Although our model is only applicable to $pA$ collisions, the feature of the heavy flavor $v_2$ shown in Fig.~\ref{fig:v2-d0} is also qualitatively in agreement with the recent ATLAS measurement on the elliptic flow of muons from the decay of charm and bottom hadrons measured in $pp$ collisions\cite{Aad:2019aol}, which indicates that the bottom flow is suppressed as compared to the charm flow.

\begin{figure}[!h]
\includegraphics[width=0.45\textwidth]{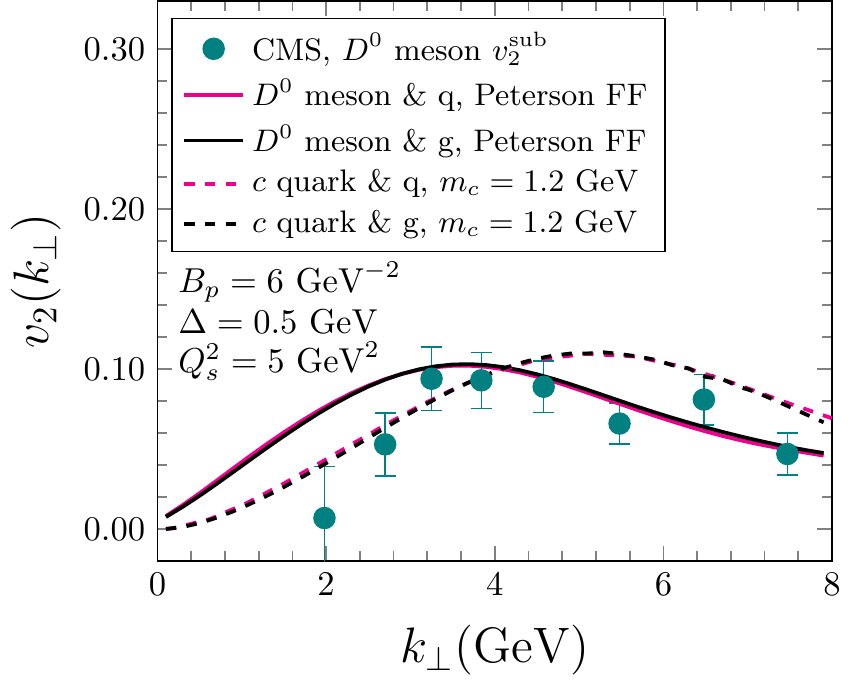}
\caption{Elliptic flows of open heavy mesons in the $g\to \Q\Qb+q$ and $g\to \Q\Qb+g$ channels.}
\label{fig:v2-d0-channels}
\end{figure}

At last, we have also numerically evaluated the elliptic flow of open heavy mesons in the $g \to \Q\Qb+g$ channel with an incoming gluon from the proton projectile as the reference. As shown in Fig. \ref{fig:v2-d0-channels}, we find again that there is little numerical difference between these two channels, which means that the elliptic flow coefficients are insensitive to the type of reference particle. This conclusion is the same as that for the ellipitic flow of heavy quarkonium. The direct comparison with the experimental data on the correlation should involve a proper combination of two channels. However, we expect no significant deviations from the contribution of each individual channel. We have also tested that our numerical result does not depend strongly on the fragmentation function of the reference parton, again due to the cancellation in the $v_2$ calculation.

\section{Conclusion}
In summary, we have studied the azimuthal angle correlation, and derived the analytic expressions of the second Fourier harmonic coefficients, for heavy quarkonium, heavy quarks or heavy mesons with respect to the reference quark, in the dilute-dense factorization in pA collisions. Our calculations of the elliptic flow of heavy mesons ($J/\psi$ and $D^0$ mesons) in the CGC formalism are consistent with the data from the CMS collaboration. In addition, we made predictions for the elliptic flow of $\Upsilon$ and $B$ mesons, which could be measured in the future. As explained above, we predict that the $v_2$ of $B$ mesons should be significantly suppressed as compared to that of the $D$ meson due to the mass dependence in the open heavy flavor channels, although we find little mass dependence in the heavy quarkonium channels. 
 
To explore robustness of the collectivity of heavy mesons due to initial state effects in the CGC formalism, we computed the corresponding $v_2$ in several slightly different setups of approximations and model inputs. First, we computed the heavy quarkonium $v_2$ by integrating over the relative momentum of the heavy quark pair $\Q\Qb$ analytically from $0$ to infinity as an approximation in Ref.~\cite{Zhang:2019dth}. In the meantime, we can also calculate the correlation by numerically integrating the relative momentum of the heavy quark pair $\Q\Qb$ within the kinematical range according to the CEM. These two calculations yield similar numerical results for the correlations. Furthermore, we find that the resulting correlation remains almost the same if we use a gluon or a charged hadron as the reference particle instead of a quark as we originally chose. In addition, little dependence on various types of heavy meson fragmentation functions was found in our numerical calculations.

Future experimental and theoretical efforts along this line may be able to help us explore the origin of collectivity in high-multiplicity events in high-energy $pA$ collisions. The collectivity of heavy mesons could also provide us an interesting gateway to understand the properties of the initial-state dense gluon matter inside high-energy hadrons.

\textit{Note added:} During the completion of this manuscript, we have made the prediction for the  $v_2$ of the $B$ meson available to the CMS collaboration. After convoluting with the decay kinematics for the $B\to D^0$ decay by using PYTHIA \cite{Sjostrand:2007gs}, CMS collaboration finds our calculation with only initial state effects to be consistent with their recent non-prompt $D$ meson $v_2$ data\cite{CMS:2019isc}.

\begin{acknowledgments}
We thank Zhenyu Chen, Wei Li, and Feng Yuan for useful discussions and comments. This material is partly supported by the Natural Science Foundation of China (NSFC) under Grant Nos.~11575070, 11775095, 11890711, 11935007, and by the China Scholarship Council (CSC) under Grant No.~201906775042. CM and SYW are supported by the Agence Nationale de la Recherche under the project ANR-16-CE31-0019-02. 

We have employed Jaxodraw \cite{Binosi:2003yf,Binosi:2008ig} to draw the Feynman diagrams in this paper.
\end{acknowledgments}

\appendix

\begin{widetext}

\section{Detailed derivation of the correlations between heavy mesons and the reference quark}
\label{ch:kappa2-d0}

By substituting Eq. (\ref{eq:cs-d0}) into Eq. (\ref{eq:kappan}), we get
\begin{align}
\frac{d\kappa_2}{dy_{D}dk_{D \perp}}
= ~ &
k_{D\perp} \mathcal{N} \int d\phi_{k_{D\perp}} 
\int dy_q d^2k_{q\perp} e^{-i2(\phi_{k_{D\perp}} - \phi_{k_{q\perp}})} 
\int \frac{d^2 b_1 d^2b_2 d^2 r_1 d^2 r_2 d^2 r}{(2\pi)^7} 
\int \frac{d\zeta}{\zeta^2} \int dz \frac{D(z)}{z^2}
e^{-i \frac{\vec k_{D\perp}}{z} \cdot \frac{\vec r_1}{\zeta}} e^{-i \vec k_{q\perp} \cdot \vec r_2}
\nonumber \\
& \times 
W(x_g,x_q, b_1, b_2,r_1,r_2) 
k_g^+ \sum_{\alpha\beta}^{\lambda} \psi_{\alpha\beta}^{T\lambda}(\vec r) \psi_{\alpha\beta}^{T\lambda*} (\vec r'=\vec r-\frac{\vec r_1}{\zeta}) \langle DDD \rangle|_{\vec r'=\vec r-\frac{\vec r_1}{\zeta}}.
\end{align}
The leading-$N_c$ terms of the scattering amplitude, $\langle DDD \rangle$, do not contribute to the second harmonic. Therefore, in this subsection, we only keep the following non-zero contributions,
\begin{align}
\langle DDD \rangle|_{\vec r'=\vec r-\vec r_1/\zeta} =
\frac{Q_s^4}{4N_c^2} \int d\xi \int_0^\xi d\eta 
\Bigl[
\langle DDD \rangle_1 + \langle DDD \rangle_2 - \langle DDD \rangle_3 -\langle DDD \rangle_4
\Bigr],
\end{align}
where,
\begin{align}
\langle DDD \rangle_1 
= 
~&
\exp \left[ -\frac{Q_s^2}{4}(\frac{r_1^2}{\zeta^2} + r_2^2) \right]
\exp \left[ -\frac{\eta Q_s^2}{2} (\vec b_1 - \vec b_2 + (1-\zeta)\vec r - \frac{1-\zeta}{2\zeta} \vec r_1)^2 \right]
\nonumber \\
&
\exp \left[ \frac{\eta Q_s^2}{8} (\frac{\vec r_1}{\zeta} - \vec r_2)^2 \right]
(\frac{\vec r_1}{\zeta} \cdot \vec r_2)^2,
\label{eq:ddd1}
\\
\langle DDD \rangle_2 
= 
~&
\exp \left[ -\frac{Q_s^2}{4} (2r_1^2+r_2^2) \right]
\exp \left[ -\frac{\eta Q_s^2}{2} [(\vec b_1-\vec b_2)^2- \frac{r_1^2+r_2^2}{4}] \right]
\exp \left[ -\frac{\eta Q_s^2 \vec r_1 \cdot \vec r_2}{4} 
\right]
2 (\vec r_1 \cdot \vec r_2)^2,
\label{eq:ddd2}
\\
\langle DDD \rangle_3 
= 
~&
\exp \left[ -\frac{Q_s^2}{4} (2r_1^2 + r_2^2 + (\zeta^2+(1-\zeta)^2) r^2 + 2 (1-2\zeta) \vec r \cdot \vec r_1) \right]
\nonumber \\
&
\exp \left[ -\frac{\eta Q_s^2}{2} [(\vec b_1 - \vec b_2 + \frac{1-\zeta}{2} \vec r)^2 - \frac{r_2^2}{4} -  \frac{(\vec r_1 +(1-\zeta)\vec r)^2}{4}] \right] 
\nonumber \\
&
\exp \left[- \frac{\eta Q_s^2}{4} (\vec r_1\cdot \vec r_2 + (1-\zeta) \vec r \cdot \vec r_2) \right]
(\vec r_1 \cdot \vec r_2 + (1-\zeta)\vec r \cdot \vec r_2)^2 
\nonumber \\
+ 
&
\exp \left[ -\frac{Q_s^2}{4} (2r_1^2 + r_2^2 + (\zeta^2+(1-\zeta)^2) r^2) + 2(1-2\zeta) \vec r \cdot \vec r_1 \right]
\nonumber \\
&
\exp \left[ -\frac{\eta Q_s^2}{2} [(\vec b_1 - \vec b_2 - \frac{\zeta}{2}\vec r)^2 - \frac{r_2^2}{4} - \frac{(\vec r_1 - \zeta \vec r)^2}{4}] \right] 
\nonumber \\
&
\exp \left[\frac{\eta Q_s^2}{4} (\vec r_1\cdot \vec r_2 - \zeta \vec r \cdot \vec r_2) \right]
(\vec r_1 \cdot \vec r_2 - \zeta \vec r \cdot \vec r_2)^2 ,
\label{eq:ddd3}
\\
\langle DDD \rangle_4 
= 
~&
\exp \left[ -\frac{Q_s^2}{4} (\frac{r_1^2}{\zeta^2} + r_2^2 + (\zeta^2+(1-\zeta)^2) r^2 - \frac{2(1-\zeta)}{\zeta} \vec r_1 \cdot \vec r) \right]
\nonumber \\
&
\exp \left[ -\frac{\eta Q_s^2}{2} [(\vec b_1 -\vec b_2+ \frac{\vec r_1}{2} - \frac{\zeta}{2} \vec r)^2 - \frac{r_2^2}{4} -\frac{\zeta^2 r^2}{4}] \right]
\nonumber \\
& 
\exp \left[ \frac{\eta Q_s^2}{4} \zeta \vec r \cdot \vec r_2 \right]
\zeta^2 (\vec r \cdot \vec r_2)^2
\nonumber \\
+
&
\exp \left[ -\frac{Q_s^2}{4} (\frac{r_1^2}{\zeta^2} + r_2^2 + (\zeta^2 + (1-\zeta)^2)r^2 - \frac{2(1-\zeta)}{\zeta} \vec r_1 \cdot \vec r) \right]
\nonumber \\
&
\exp \left[ -\frac{\eta Q_s^2}{2} [(\vec b_1 - \vec b_2 - \frac{(1-\zeta) \vec r_1}{2\zeta} + \frac{(1-\zeta) \vec r}{2})^2 - \frac{r_2^2}{4} - (\frac{\vec r_1}{2\zeta} -\frac{(1-\zeta) \vec r}{2})^2] \right]
\nonumber \\
& 
\exp \left[ -\frac{\eta Q_s^2}{2} (\frac{1}{2\zeta} \vec r_1 \cdot \vec r_2 - \frac{(1-\zeta) \vec r_2 \cdot \vec r}{2}) \right]
(\frac{1}{\zeta}\vec r_1 \cdot \vec r_2 - (1-\zeta) \vec r_2 \cdot \vec r)^2.
\label{eq:ddd4}
\end{align}

The $b_{1,2}$ dependece in Eqs. (\ref{eq:ddd1}-\ref{eq:ddd1}) always takes the form $\exp[-\frac{\eta Q_s^2}{2}(\vec b_1-\vec b_2-\vec X)^2]$, where $\vec X$ could be any two-dimensional vector. 

Utilizing the following relation
\begin{align}
\int \frac{d^2b_1 d^2b_2}{(2\pi)^2} \frac{1}{\pi^2 B_p^2} e^{-\frac{b_1^2+b_2^2}{B_p}} e^{-\frac{\eta Q_s^2}{2}(\vec b_1- \vec b_2 - \vec X)^2} 
= 
\frac{1}{4\pi^2} \frac{1}{1+\eta Q_s^2 B_p} 
\exp \left[ - \frac{\eta Q_s^2 X^2}{2(1+\eta Q_s^2 B_p)} \right],
\end{align}
the differential $\kappa_2$ becomes,
\begin{align}
\frac{d\kappa_2}{dy_{D}dk_{D \perp}}
= ~ &
k_{D\perp} \mathcal{N} \int d\phi_{k_{D\perp}} 
\int dy_q d^2k_{q\perp} e^{-i2(\phi_{k_{D\perp}} - \phi_{k_{q\perp}})} 
\int \frac{d^2 b_1 d^2b_2 d^2 r_1 d^2 r_2 d^2 r}{(2\pi)^7} 
\int \frac{d\zeta}{\zeta^2} \int dz \frac{D(z)}{z^2}
e^{-i \frac{\vec k_{D\perp}}{z} \cdot \frac{\vec r_1}{\zeta}} e^{-i \vec k_{q\perp} \cdot \vec r_2}
\nonumber \\
& \times 
x_g f_g(x_g) x_q f_q (x_q) \exp \left[ -\frac{\Delta^2 (r_1^2 + r_2^2)}{4} \right]
k_g^+ \sum_{\alpha\beta}^{\lambda} \psi_{\alpha\beta}^{T\lambda}(\vec r) \psi_{\alpha\beta}^{T\lambda*} (\vec r'=\vec r-\frac{\vec r_1}{\zeta}) 
\nonumber \\
& \times
\frac{Q_s^4}{4N_c^2} \int d\xi \int_0^\xi d\eta
\left[\int \langle DDD\rangle_1+\int \langle DDD\rangle_2-\int \langle DDD\rangle_3-\int \langle DDD\rangle_4\right],
\end{align}
where, $\int \langle DDD \rangle_i $ is defined as
\begin{align}
\int \langle DDD \rangle_i = 
\int \frac{d^2b_1d^2b_2}{(2\pi)^2} \frac{1}{\pi^2 B_p^2} e^{-\frac{b_1^2+b_2^2}{B_p}} 
\langle DDD \rangle_i,
\end{align}
with
\begin{align}
\int\langle DDD \rangle_1
= 
~&
\exp\left[ - \frac{\eta Q_s^2}{2(1+ \eta Q_s^2 B_p)} ((1-\zeta)\vec r - \frac{1-\zeta}{2\zeta} \vec r_1)^2 \right]
\exp \left[ -\frac{Q_s^2}{4}(\frac{r_1^2}{\zeta^2} + r_2^2) \right]
\nonumber\\
&
\exp \left[ \frac{\eta Q_s^2}{8} (\frac{\vec r_1}{\zeta} - \vec r_2)^2 \right]
(\frac{\vec r_1}{\zeta} \cdot \vec r_2)^2 ,
\\
\int \langle DDD \rangle_2 
= 
~&
\exp \left[ -\frac{Q_s^2}{4} (2r_1^2+r_2^2) \right]
\exp \left[ \frac{\eta Q_s^2}{8} (r_1^2+r_2^2) \right]
\exp \left[ -\frac{\eta Q_s^2 \vec r_1 \cdot \vec r_2}{4} 
\right]
2 (\vec r_1 \cdot \vec r_2)^2,
\\
\int \langle DDD \rangle_3 
= 
~&
\exp \left[ -\frac{Q_s^2}{4} (2r_1^2 + r_2^2 + (\zeta^2+(1-\zeta)^2) r^2 + 2 (1-2\zeta) \vec r \cdot \vec r_1) \right]
\nonumber\\
&
\exp \left[ -\frac{\eta Q_s^2}{8(1+\eta Q_s^2 B_p)} (1-\zeta)^2 r^2 \right]
\exp \left[ \frac{\eta Q_s^2}{8} [ r_2^2  + (\vec r_1 +(1-\zeta)\vec r)^2] \right] 
\nonumber\\
&
\exp \left[- \frac{\eta Q_s^2}{4} (\vec r_1\cdot \vec r_2 + (1-\zeta) \vec r \cdot \vec r_2) \right]
(\vec r_1 \cdot \vec r_2 + (1-\zeta)\vec r \cdot \vec r_2)^2 
\\
+ 
&
\exp \left[ -\frac{Q_s^2}{4} (2r_1^2 + r_2^2 + (\zeta^2+(1-\zeta)^2) r^2) + 2(1-2\zeta) \vec r \cdot \vec r_1 \right]
\nonumber\\
&
\exp \left[ -\frac{\eta Q_s^2}{8(1+\eta Q_s^2 B_p)} \zeta^2 r^2  \right]
\exp \left[ \frac{\eta Q_s^2}{8} [ r_2^2 + (\vec r_1 - \zeta \vec r)^2 ] \right] 
\nonumber\\
&
\exp \left[\frac{\eta Q_s^2}{4} (\vec r_1\cdot \vec r_2 - \zeta \vec r \cdot \vec r_2) \right]
(\vec r_1 \cdot \vec r_2 - \zeta \vec r \cdot \vec r_2)^2 ,
\\
\int \langle DDD \rangle_4 
= 
~&
\exp \left[ -\frac{Q_s^2}{4} (\frac{r_1^2}{\zeta^2} + r_2^2 + (\zeta^2+(1-\zeta)^2) r^2 - \frac{2(1-\zeta)}{\zeta} \vec r_1 \cdot \vec r) \right]
\nonumber\\
&
\exp \left[-\frac{\eta Q_s^2}{8(1+\eta Q_s^2 B_p)} (\vec r_1-\zeta \vec r)^2 \right]
\exp \left[\frac{\eta Q_s^2}{8} [ r_2^2 + \zeta^2 r^2] \right]
\nonumber\\
& 
\exp \left[ \frac{\eta Q_s^2}{4} \zeta \vec r \cdot \vec r_2 \right]
\zeta^2 (\vec r \cdot \vec r_2)^2
\nonumber\\
+
&
\exp \left[ -\frac{Q_s^2}{4} (\frac{r_1^2}{\zeta^2} + r_2^2 + (\zeta^2 + (1-\zeta)^2)r^2 - \frac{2(1-\zeta)}{\zeta} \vec r_1 \cdot \vec r) \right]
\nonumber\\
&
\exp \left[-\frac{\eta Q_s^2 (1-\zeta)^2}{8(1+\eta Q_s^2 B_p)} (\vec r-\frac{\vec r_1}{\zeta})^2 \right]
\exp \left[\frac{\eta Q_s^2}{8} [r_2^2 + (\frac{\vec r_1}{\zeta} -(1-\zeta)\vec r)^2] \right]
\nonumber\\
& 
\exp \left[ -\frac{\eta Q_s^2}{4} (\frac{\vec r_1}{\zeta} \cdot \vec r_2 - (1-\zeta) \vec r_2 \cdot \vec r ) \right]
(\frac{1}{\zeta}\vec r_1 \cdot \vec r_2 - (1-\zeta) \vec r_2 \cdot \vec r)^2.
\end{align}

Furthermore, we can remove the integrals over the azimuthal angles of transverse momenta using the following relation
\begin{align}
\int \frac{d\phi_{k_{D\perp}} d\phi_{k_{q\perp}}}{(2\pi)^2} e^{-i 2(\phi_{k_{D\perp}}-\phi_{k_{q\perp}})} 
e^{-i \frac{\vec k_{D\perp}}{z} \cdot \frac{\vec r_1}{\zeta}} e^{-i \vec k_{q\perp} \cdot \vec r_2}
= J_2 (|\frac{\vec k_{D\perp}}{z}| |\frac{\vec r_1}{\zeta}|) J_2(|\vec k_{q\perp}||\vec r_2|) \cos [2(\phi_{r_1} - \phi_{r_2})].
\end{align}

To compute the $d^2r_2$ integral analytically, we can perform a Taylor expansion of the exponentials that contain the factor $\vec r_2 \cdot \vec Y$, where $\vec Y$ could be any vector. The $|\vec r_2|$ and $\phi_{r_2}$ dependences are now separated. Using the following relations
\begin{align}
& 
\int_0^{2\pi} \frac{d\phi_{r_2}}{2\pi} \cos (2\phi_{r_2}) \cos^{2m+2} \phi_{r_2} 
= \frac{1}{2^{2m+2}}\frac{(2m+2)!}{m! (m+2)!},
\\
& 
\int_0^{2\pi} \frac{d\phi_{r_2}}{2\pi} \cos (2\phi_{r_2}) \cos^{2m+1} \phi_{r_2} 
= 0,
\\
& 
\int_0^{2\pi} \frac{d\phi_{r_2}}{2\pi} \sin (2\phi_{r_2}) \cos^{l+2} \phi_{r_2} 
= 0,
\\
& 
\int r_2 dr_2 J_2 (|\vec k_{q\perp}| |\vec r_2|) \exp\left[ - a_q r_2^2 \right] r_2^{2m+2} = (m+2)! \frac{k_{q\perp}^2}{16} \frac{1}{a_q^{m+3}} {}_1F_1 (m+3, 3, -\frac{k_{q\perp}^2}{4a_q}),
\end{align}
we can obtain Eq. (\ref{eq:kappa2-d0}). 

\end{widetext}

\end{document}